\documentclass[11pt,twoside]{article}

\usepackage{asp2006}
\usepackage{epsf}
\usepackage{epsfig}
\usepackage{lscape}

\begin{document}
\newcommand{\kms}{km~s$^{-1}$}
\newcommand{\Msun}{M_{\odot}}
\newcommand{\Lsun}{L_{\odot}}
\newcommand{\ML}{M_{\odot}/L_{\odot}}
\newcommand{\etal}{{et al.}\ }
\newcommand{\hhh}{h_{100}}
\newcommand{\hsq}{h_{100}^{-2}}
\newcommand{\tn}{\tablenotemark}
\newcommand{\mdot}{\dot{M}}
\newcommand{\p}{^\prime}
\newcommand{\kmsMpc}{km~s$^{-1}$~Mpc$^{-1}$}

\title{Our CMB motion: The Role of the Local Void}   
\author{R. Brent Tully}   
\affil{Institute for Astronomy, University of Hawaii}    

\begin{abstract} 
A database with a high density of accurate distances is used to investigate the contributions to the motion of our Galaxy.  It is found that the motion of the Local Group separates remarkably cleanly into 3 components: a large-scale attraction toward structure in the `Great Attractor' sector, a mid-scale attraction toward the Virgo Cluster, and a local-scale `repulsion' from the Local Void.  These 3 components cause motions of comparable amplitudes and, conveniently, are directed almost orthogonal to one another.   
\end{abstract}


\section{Introduction}   
The dipole temperature variation of the cosmic microwave background (CMB) indicates that we have a peculiar motion of over 600~\kms\ in the Local Group rest frame.  Part of that motion is directed toward the concentration of galaxies in the vicinity of the Virgo Cluster  \citep{1982ApJ...258...64A, 1984ApJ...281...31T, 2000ApJ...530..625T}.  Another part is directed toward the structures in the Norma--Hydra--Centaurus regions  \citep{1988ApJ...326...19L}  with a possible relevance of the background Shapley Concentration  \citep{1989Natur.338..562S}.
These sources strongly influence our motion but there is yet another very substantial component.
There have been hints of the importance of this  component that have given rise to discussion of the `local velocity anomaly'  \citep{1988lsmu.book..115F, 1988lsmu.book..169T, 1992ApJS...80..479T}.  Today there are many more excellent distances to nearby galaxies and these new distances clarify the situation considerably.  Remarkably, it is possible to separate the influences on our motion on scales beyond the Local Group into three cleanly separated components.

The most important of the new distance information is coming from Hubble Space Telescope (HST) imaging of nearby galaxies of all types and the color-magnitude diagrams of the resolved stars that result from those observations.  A single orbit sequence of observations with Advanced Camera for Surveys (ACS) provides sufficient definition of the tip of the red giant branch (TRGB) for the determination of a distance with relative accuracy of 5\% (without getting into zero-point issues) for any galaxy within 10~Mpc     \citep{2004AJ....127.2031K, 2006AJ....131.1361K, 2006astro.ph..3073M}.  Presently, HST has provided information for over 200 galaxies within this distance, roughly half of what we might hope ultimately to obtain.

The TRGB measurements provide outstanding distances to nearby galaxies.   At larger distances, the present analysis relies on the following sources: the HST Key Project   \citep{2001ApJ...553...47F}  for Cepheid distances and the zero-point to the distance scale that the Key Project provides, the Surface Brightness Fluctuation (SBF) measures of  \citet{2001ApJ...546..681T}, and a personal compilation of luminosity--linewidth measures    \citep{1977A&A....54..661T}  extending on the sample discussed by  \citep{2000ApJ...533..744T}.  This compendium of distances is restricted to $V \sim 3000$~\kms, including the Centaurus and Hydra clusters at it's limit.  The catalog contains distance estimates for 1485 galaxies in 661 groups.  The catalog will be made available on-line with the publication of an article that describes the present results more completely.  

\section{The Peculiar Velocity Field within 3000~\kms}

\begin{figure}[]
\begin{center}
\includegraphics[scale=0.6]{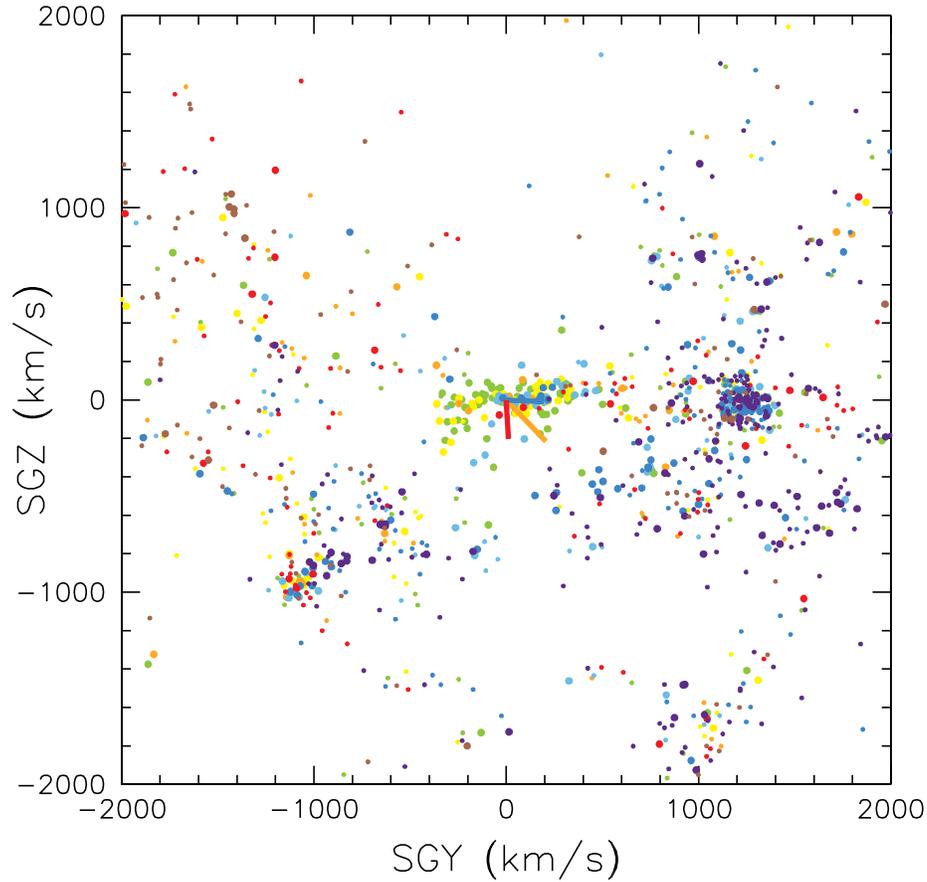}
\caption{Motion within the Local Supercluster in the rest frame of the Local Group.   Galaxies with measured distances are located at their positions in supergalactic coordinates SGY, SGZ with our Galaxy at the origin. Depth $\pm 2000$~\kms\ in SGX. Peculiar velocities are calculated assuming H$_0 = 75$~\kmsMpc\ and values are indicated with the following color code:   {\it purple} $V_{pec} \leq -400$~\kms; {\it dark blue} $-400 < V_{pec} \leq -200$~\kms; {\it light blue} $-200 < V_{pec} \leq -100$~\kms; {\it green} $-100 < V_{pec} < 0$~\kms; {\it yellow} $0 \leq V_{pec} < 100$~\kms; {\it orange} $100 \leq V_{pec} < 200$~\kms; {\it red} $200 \leq V_{pec} < 400$~\kms; {\it brown} $400 \leq V_{pec}$~\kms.  Large Symbols: distances from Cepheid, TRGB, or SBF methods; small symbols: luminosity--linewidth distances.  The vectors emanating from our position at the origin indicate our motion relative to these galaxies.  The amplitude of these vectors is scaled in accordance with the scale in \kms\ of the axes.}
\label{lsc_vectors_big}
\end{center}
\end{figure}

Figure 1 illustrates the peculiar velocities of the galaxies within an inner box of $\pm 2000$~\kms\ of our position.  Several aspects of the plot are striking.  The obvious cluster at SGY=1200, SGZ=0 is the Virgo Cluster and there is a great predominance of blue shades among the galaxies found there.  There is an
implied motion of approach and we can presume that {\it we are being pulled toward the Virgo Cluster} rather than the other way round.  Second, we see that all the galaxies in our immediate vicinity have modest relative velocities, causing them to be colored green or yellow.  {\it The filament containing our Galaxy and all our immediate neighbors is heading somewhere as a unit, with little internal velocity dispersion.}    Third, there is a tendency for objects between one o'clock and 8 o'clock in Fig.~1 to be
in shades of blue and objects in the opposite sector to be in shades of red.  {\it We are headed toward 4:30.}  Indeed, the vector of our motion with respect to our full sample of distance estimates  has an
amplitude of 298~\kms\ toward supergalactic coordinates $L=87, B=-46$ (Galactic coordinates $\ell=225, b= +38$).  This vector is indicated in orange in Fig.~1.

The Virgo Cluster must be causing part of this motion.  A recent analysis that gives attention to the infall pattern in the immediate vicinity of the Virgo Cluster  \citep{2005ApJ...635L.113M}  ascribes a mass approaching $1 \times 10^{15} \Msun$ to the cluster within a radius of 2~Mpc.  It follows that we have a peculiar motion of $\sim 200$~\kms\ toward Virgo at our location.  If a vector of this amplitude and direction (the blue vector in Fig.~1) is subtracted from the orange vector then the result is the red vector, with amplitude 211~\kms\ toward $L=12, B=-75$ ($\ell=213, b=-3$).   This vector is not very sensitive to the details of the vector subtraction of the Virgo component because the red and blue vectors are almost orthogonal; a variation of $\pm 50$~\kms\ toward Virgo affects the amplitude of the red vector by only 10~\kms\ and the direction by $15^{\circ}$. 

\section{The Local Void}

What are we to make of this large motion orthogonal to the plane of the local filament?  Consider Figure~2.   It is reaffirmed that {\it all our neighbors within our filament are traveling with us in this downward excursion.}  Second, this motion is a {\bf local} phenomenon: the filament below us at SGZ=$-$500~\kms, the Leo Spur in the Nearby Galaxies Atlas,  \citep{1987nga..book.....T} is not participating in this activity.  To the contrary, after subtracting off our motion {\it the Leo Spur has a net motion toward positive SGZ!}  This motion is well established by the accurate TRGB distance measures reflected in the black peculiar velocity vectors shown in Fig.~2. 

There is manifestly no source of attraction that would pull us down toward negative SGZ while pulling the Leo Spur up.   It is to be concluded that our motion is not toward anything but away from the Local Void. (The motion of the Leo Spur might in part be due to a push from a void below it and in part the pull of the equatorial plane of the Local Supercluster.) The Local Void  \citep{1987nga..book.....T}  is impressively extensive and impressively near.  The void lurks at the edge of the Local Group.  It actually seems to have two components; a smaller void with a long dimension of $\sim 35$~Mpc enclosed within a larger void with a long dimension of $\sim 60$~Mpc.  We lie on a wall that is shared by the two, which we are calling the Inner and Greater Local Voids.  If one wants to study the properties of voids, one does not have to look very far.  

The Local Void is not a total void.  A good TRGB distance and HI velocity exists for the lonely dwarf galaxy ESO~461$-$36 \citep{2006AJ....131.1361K}.  This galaxy has the peculiar velocity $-216$~\kms\ indicated by the vector in Fig.~2.    This galaxy is chasing us out of the void.  It's motion is additive to the orange vector of Fig.~1 so it's motion with respect to the rest frame given by our $\pm 3000$~\kms\ sample is over 500~\kms.
\begin{figure}[]
\begin{center}
\includegraphics[scale=0.45]{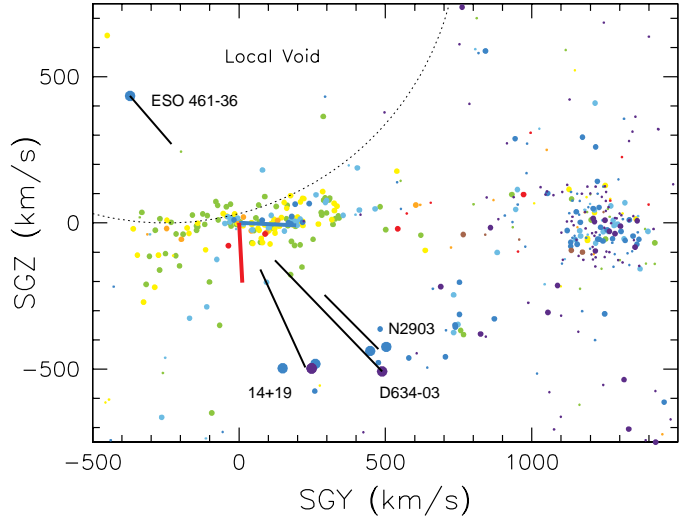}
\caption{Motion away from the Local Void.  The dotted curve roughly describes the surface of the void in our vicinity.  ESO~461-36 is an isolated dwarf galaxy within the Local Void with a high velocity toward the boundary of the void.  The black vectors illustrate the peculiar velocities of selected galaxies and groups in the reference frame of the Local Group while the blue and red vectors illustrate the dissection of the peculiar velocity of the Local Group in the reference frame of the Local Supercluster into components toward the Virgo Cluster and away from the Local Void.}
\label{lvdw}
\end{center}
\end{figure}

\section{The Large-Scale Component of our Peculiar Velocity}

\begin{figure}
\begin{center}
\includegraphics[scale=0.5]{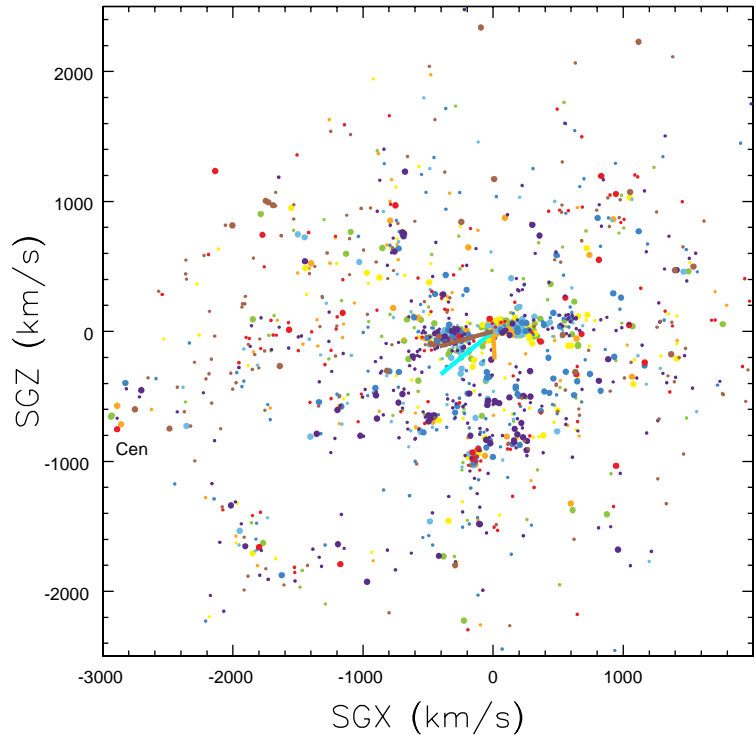}
\caption{Decomposition of the vectors of the motion of the Local Group.  Peculiar velocities of galaxies with observed distances are shown with the same color code introduced in Fig.~1.  The orange vector indicates the motion of the Local Group with respect to this sample.  The cyan vector indicates the motion of the Local Group with respect to the rest frame established by the CMB.  The brown vector is the vector difference between these two and is attributed to attractors on scales greater than 3000~\kms.}
\label{allv}
\end{center}
\end{figure}

The motion defined by our sample of galaxies with distances within 3000~\kms\  does not agree with the global motion inferred from the CMB dipole anisotropy.  Figure~3 shows these two vectors (orange and cyan respectively) and the resultant vector (brown) after the local contribution is subtracted from the CMB  cumulative motion.  This resultant vector is taken to be generated by forces on larger scales than sampled by our distance catalog.  The cyan global CMB vector is taken   \citep{1993ApJ...419....1K}  to have an amplitude in the Local Group rest frame of 619~\kms\  directed toward $L=140.2, B=-32.2$ ($\ell=275.2, b=+28.7$).  Subtracting the orange Local Supercluster vector leaves the brown large-scale vector of 446~\kms\ toward $L=163, B=-15$ ($\ell=300, b=+15$).  

The directions of these various vectors are displayed on the Aitoff projection of Figure~4 along with the positions of features that might be considered relevant.    Partially but not completely by chance, the
three components of the CMB decomposition are almost orthogonal to each other, with Virgo near +SGY, the Local Void reflex near $-$SGZ, and the large scale (Cen/Shapley) component near $-$SGX
(it is no coincidence that the attractors are at the Supergalactic equator and the repulser is at the Supergalactic pole).  Two recently determined dipoles in the distribution of galaxies are plotted in Fig.~4. One is the dipole in the distribution of 2MASS selected galaxies     \citep{2006MNRAS.368.1515E}  while the other is the dipole in the distribution of X-ray selected rich clusters  \citep{2006ApJ...645.1043K}.  The former is drawn from a sample that peaks in number at $\sim 4000$~\kms\ and provides poor coverage beyond 10,000~\kms\ while the latter is only becoming interesting beyond 4000~\kms.  Interestingly, the 2MASS dipole is offset from the CMB dipole in the direction of the orange velocity vector defined by our 3,000~\kms\ sample while the X-ray cluster dipole is offset from the CMB dipole in the direction of the brown large scale residual velocity vector.   The 2MASS sample is relatively local, strongly influenced by galaxies within 3000~\kms\ that we are sampling with our distance catalog.  The X-ray selected
sample has very little overlap with our domain and provides a degree of confirmation of our decomposition of the CMB vector into local and large scale components.  It follows that models that try to describe the large scale motion need only explain the reduced amplitude of the residual large scale component and the direction of this residual component which agrees reasonably with the direction of obvious concentrations in the distribution of galaxies. 
\begin{figure}[]
\begin{center}
\includegraphics[scale=0.7]{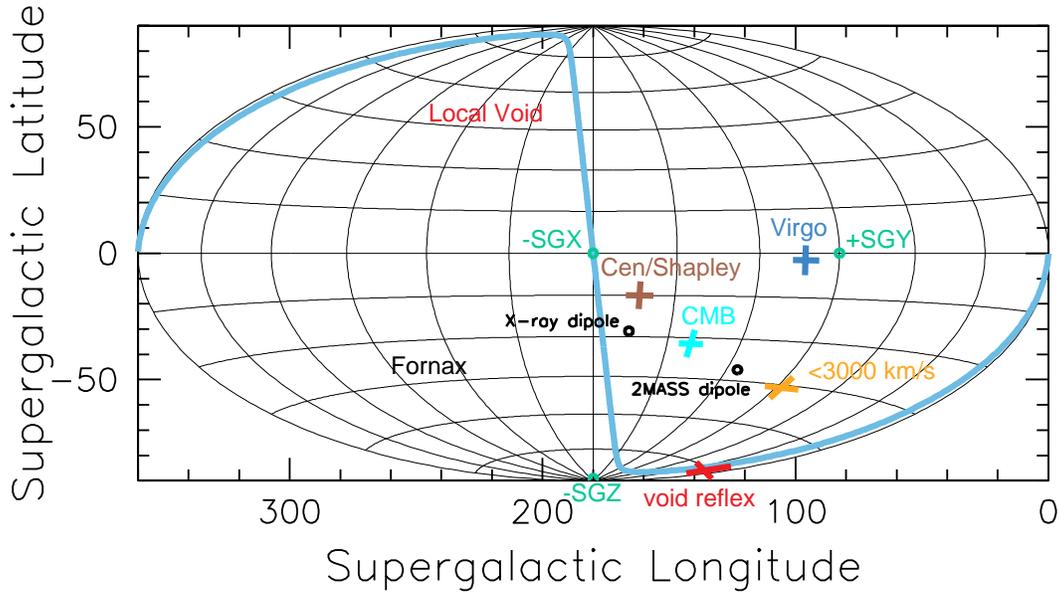}
\caption{Components of the motion of the Local Group projected onto the sky.  The motion of 619~\kms\ given by the CMB dipole in the direction indicated by the cyan cross can be decomposed into the 
298~\kms\ component defined by the distance measures discussed in this paper, confined to the traditional Local Supercluster, and labeled $<3000$~\kms\ in orange and the 446~\kms\ residual located by the brown cross that can be ascribed to large scale structures in Norma--Hydra--Centaurus and to the background.  The  Local Supercluster $<3000$~\kms\ component can in turn be separated into a component of 200~\kms\ toward the Virgo Cluster, at the blue cross, and a component of 211~\kms\ away from the Local Void, at the (distorted) red cross.  The Local Void reflex, Virgo, and large scale Cen/Shapley attractions are almost orthogonal to one another, toward the --SGZ, +SGY, and --SGX axes respectively.  The X-ray dipole direction is found to lie close to the direction of the large scale Cen/Shapley vector which can be understood since the characteristic distances of the X-ray cluster sample are large.  By contrast the 2MASS dipole lies midway between the CMB and $<3000$~\kms\ Local Supercluster vector directions.  It is inferred that the 2MASS  dipole is determined relatively locally.}
\label{aitoffv}
\end{center}
\end{figure}

\section{Summary}

1. A sample of 1485 distance measure for galaxies with velocities within 3000 \kms\  reveal that the Local Group has a motion with respect to this sample of 298~\kms\ in a direction with components toward the Virgo Cluster and the supergalactic south pole.

\noindent
2. The galaxies within our local filament share this motion with very low internal random motions.

\noindent
3. The motion toward the supergalactic south pole is in the sense of an evacuation of the Local Void which lies north of the supergalactic equator just beyond the Local Group.  A lonely dwarf galaxy in the void is seen to be moving toward the edge of the void with a velocity of over 500~\kms\ with respect to the galaxies of our sample.

\noindent
4. The Leo Spur, the filament that lies below the supergalactic plane, has a motion of several hundred \kms\ {\it upward} with respect to the galaxies in our sample.  In co-moving coordinates our filament and the Leo Spur are converging although in expanding coordinates the two structures are almost stationary.

\noindent
5. If the Local Group motion with respect to our sample limited to 3000~\kms\ is subtracted from the CMB dipole motion, the resultant motion of 446~\kms\ to be attributed to the large scale distribution of matter is found to be reasonably aligned with the observed overdensity of galaxies in the general direction of Centaurus.  

\acknowledgements 

An expanded version of this talk is being prepared for a refereed publication.  I thank my colleagues in that enterprise: Helen Courtois, Igor Karachentsev, Dale Kocevski, Alan Peel, Luca Rizzi, and Ed Shaya.   This research is supported by HST awards GO-9162, 10210 and 10905, a SIM Key Project, and NSF award AST-0307706.



\bibliographystyle{asp}
\bibliography{paper}

\begin{thebibliography}{}
\expandafter\ifx\csname natexlab\endcsname\relax\def\natexlab#1{#1}\fi

\bibitem[{{Aaronson} {et~al.}(1982){Aaronson}, {Huchra}, {Mould}, {Schechter},
  \& {Tully}}]{1982ApJ...258...64A}
{Aaronson}, M., {Huchra}, J., {Mould}, J., {Schechter}, P.~L., \& {Tully},
  R.~B. 1982, \apj, 258, 64

\bibitem[{{Erdo{\u g}du} {et~al.}(2006){Erdo{\u g}du}, {Huchra}, {Lahav},
  {Colless}, {Cutri}, {Falco}, {George}, {Jarrett}, {Jones}, {Kochanek},
  {Macri}, {Mader}, {Martimbeau}, {Pahre}, {Parker}, {Rassat}, \&
  {Saunders}}]{2006MNRAS.368.1515E}
{Erdo{\u g}du}, P., {Huchra}, J.~P., {Lahav}, O., {Colless}, M., {Cutri},
  R.~M., {Falco}, E., {George}, T., {Jarrett}, T., {Jones}, D.~H., {Kochanek},
  C.~S., {Macri}, L., {Mader}, J., {Martimbeau}, N., {Pahre}, M., {Parker}, Q.,
  {Rassat}, A., \& {Saunders}, W. 2006, \mnras, 368, 1515

\bibitem[{{Faber} \& {Burstein}(1988)}]{1988lsmu.book..115F}
{Faber}, S.~M. \& {Burstein}, D. 1988, {Motions of galaxies in the neighborhood
  of the local group} (Large-Scale Motions in the Universe: A Vatican study
  Week), 115--167

\bibitem[{{Freedman} {et~al.}(2001){Freedman}, {Madore}, {Gibson}, {Ferrarese},
  {Kelson}, {Sakai}, {Mould}, {Kennicutt}, {Ford}, {Graham}, {Huchra},
  {Hughes}, {Illingworth}, {Macri}, \& {Stetson}}]{2001ApJ...553...47F}
{Freedman}, W.~L., {Madore}, B.~F., {Gibson}, B.~K., {Ferrarese}, L., {Kelson},
  D.~D., {Sakai}, S., {Mould}, J.~R., {Kennicutt}, Jr., R.~C., {Ford}, H.~C.,
  {Graham}, J.~A., {Huchra}, J.~P., {Hughes}, S.~M.~G., {Illingworth}, G.~D.,
  {Macri}, L.~M., \& {Stetson}, P.~B. 2001, \apj, 553, 47

\bibitem[{{Karachentsev} {et~al.}(2006){Karachentsev}, {Dolphin}, {Tully},
  {Sharina}, {Makarova}, {Makarov}, {Karachentseva}, {Sakai}, \&
  {Shaya}}]{2006AJ....131.1361K}
{Karachentsev}, I.~D., {Dolphin}, A., {Tully}, R.~B., {Sharina}, M.,
  {Makarova}, L., {Makarov}, D., {Karachentseva}, V., {Sakai}, S., \& {Shaya},
  E.~J. 2006, \aj, 131, 1361

\bibitem[{{Karachentsev} {et~al.}(2004){Karachentsev}, {Karachentseva},
  {Huchtmeier}, \& {Makarov}}]{2004AJ....127.2031K}
{Karachentsev}, I.~D., {Karachentseva}, V.~E., {Huchtmeier}, W.~K., \&
  {Makarov}, D.~I. 2004, \aj, 127, 2031

\bibitem[{{Kocevski} \& {Ebeling}(2006)}]{2006ApJ...645.1043K}
{Kocevski}, D.~D. \& {Ebeling}, H. 2006, \apj, 645, 1043

\bibitem[{{Kogut} {et~al.}(1993){Kogut}, {Lineweaver}, {Smoot}, {Bennett},
  {Banday}, {Boggess}, {Cheng}, {de Amici}, {Fixsen}, {Hinshaw}, {Jackson},
  {Janssen}, {Keegstra}, {Loewenstein}, {Lubin}, {Mather}, {Tenorio}, {Weiss},
  {Wilkinson}, \& {Wright}}]{1993ApJ...419....1K}
{Kogut}, A., {Lineweaver}, C., {Smoot}, G.~F., {Bennett}, C.~L., {Banday}, A.,
  {Boggess}, N.~W., {Cheng}, E.~S., {de Amici}, G., {Fixsen}, D.~J., {Hinshaw},
  G., {Jackson}, P.~D., {Janssen}, M., {Keegstra}, P., {Loewenstein}, K.,
  {Lubin}, P., {Mather}, J.~C., {Tenorio}, L., {Weiss}, R., {Wilkinson}, D.~T.,
  \& {Wright}, E.~L. 1993, \apj, 419, 1

\bibitem[{{Lynden-Bell} {et~al.}(1988){Lynden-Bell}, {Faber}, {Burstein},
  {Davies}, {Dressler}, {Terlevich}, \& {Wegner}}]{1988ApJ...326...19L}
{Lynden-Bell}, D., {Faber}, S.~M., {Burstein}, D., {Davies}, R.~L., {Dressler},
  A., {Terlevich}, R.~J., \& {Wegner}, G. 1988, \apj, 326, 19

\bibitem[{{Makarov} {et~al.}(2006){Makarov}, {Makarova}, {Rizzi}, {Tully},
  {Dolphin}, {Sakai}, \& {Shaya}}]{2006astro.ph..3073M}
{Makarov}, D., {Makarova}, L., {Rizzi}, L., {Tully}, R.~B., {Dolphin}, A.~E.,
  {Sakai}, S., \& {Shaya}, E.~J. 2006, ArXiv Astrophysics e-prints

\bibitem[{{Mohayaee} \& {Tully}(2005)}]{2005ApJ...635L.113M}
{Mohayaee}, R. \& {Tully}, R.~B. 2005, \apjl, 635, L113

\bibitem[{{Scaramella} {et~al.}(1989){Scaramella}, {Baiesi-Pillastrini},
  {Chincarini}, {Vettolani}, \& {Zamorani}}]{1989Natur.338..562S}
{Scaramella}, R., {Baiesi-Pillastrini}, G., {Chincarini}, G., {Vettolani}, G.,
  \& {Zamorani}, G. 1989, \nat, 338, 562

\bibitem[{{Tonry} {et~al.}(2000){Tonry}, {Blakeslee}, {Ajhar}, \&
  {Dressler}}]{2000ApJ...530..625T}
{Tonry}, J.~L., {Blakeslee}, J.~P., {Ajhar}, E.~A., \& {Dressler}, A. 2000,
  \apj, 530, 625

\bibitem[{{Tonry} {et~al.}(2001){Tonry}, {Dressler}, {Blakeslee}, {Ajhar},
  {Fletcher}, {Luppino}, {Metzger}, \& {Moore}}]{2001ApJ...546..681T}
{Tonry}, J.~L., {Dressler}, A., {Blakeslee}, J.~P., {Ajhar}, E.~A., {Fletcher},
  A.~B., {Luppino}, G.~A., {Metzger}, M.~R., \& {Moore}, C.~B. 2001, \apj, 546,
  681

\bibitem[{{Tully}(1988)}]{1988lsmu.book..169T}
{Tully}, R.~B. 1988, {Distances to galaxies in the field} (Large-Scale Motions
  in the Universe: A Vatican study Week), 169--177

\bibitem[{{Tully} \& {Fisher}(1977)}]{1977A&A....54..661T}
{Tully}, R.~B. \& {Fisher}, J.~R. 1977, \aap, 54, 661

\bibitem[{{Tully} \& {Fisher}(1987)}]{1987nga..book.....T}
---. 1987, {Nearby galaxies Atlas} (Cambridge: University Press, 1987)

\bibitem[{{Tully} \& {Pierce}(2000)}]{2000ApJ...533..744T}
{Tully}, R.~B. \& {Pierce}, M.~J. 2000, \apj, 533, 744

\bibitem[{{Tully} \& {Shaya}(1984)}]{1984ApJ...281...31T}
{Tully}, R.~B. \& {Shaya}, E.~J. 1984, \apj, 281, 31

\bibitem[{{Tully} {et~al.}(1992){Tully}, {Shaya}, \&
  {Pierce}}]{1992ApJS...80..479T}
{Tully}, R.~B., {Shaya}, E.~J., \& {Pierce}, M.~J. 1992, \apjs, 80, 479

\end{thebibliography}
\end{document}